\documentclass[aps,prl,twocolumn,floatfix,amsmath,showpacs]{revtex4-1}
\usepackage{natbib}
\usepackage{color}
\usepackage{graphicx}
\usepackage{amsmath}
\usepackage{amssymb}
\usepackage{amsfonts}
\DeclareGraphicsExtensions{.pdf,.png,.jpg}

\begin{document}

\title{Fractal conduction pathways governing ionic transport in a glass
}
\author{J.~L.~Iguain}
\affiliation{
Instituto de Investigaciones F\'{\i}sicas de Mar del Plata (IFIMAR) and
Departamento de F\'{\i}sica FCEyN, 
Universidad Nacional de Mar del Plata - De\'an Funes 3350, 7600 Mar del
Plata, Argentina.}
\author{F.~O.~Sanchez-Varreti}
\affiliation{ 
Universidad Tecnol\'ogica Nacional - Facultad Regional San Rafael (FRSR) – SiCo.}
\author{M.~A.~Frechero}
\affiliation{
	INQUISUR-UNS-CONICET and 
Departamento de Qu\'{\i}mica  Universidad Nacional del Sur -
Avenida Alem 1253, 8000 Bah\'{\i}a Blanca, Argentina.\\
}


\begin{abstract}
	We present a systematic characterization of the fractal conduction 
pathways governing ionic transport in a non-crystalline solid below the 
glass-transition temperature. Using classical molecular dynamics simulations 
of lithium metasilicate, we combine mobility-resolved dynamical analysis 
with a real-space description of the regions explored by lithium ions. 
Ensemble-averaged velocity autocorrelation functions rapidly decorrelate 
and do not resolve the pronounced dynamic heterogeneity of the system, 
whereas single-ion analysis reveals short-lived episodes of nearly collinear 
motion. By mapping active-site clusters over increasing time windows, 
we show that ion-conducting pathways are quasi one-dimensional at short times 
and evolve into larger, branched structures characterized by a robust fractal 
dimension $d_f\simeq1.7$. This geometry persists while the silicate backbone 
remains structurally arrested, whereas near the glass-transition temperature 
the loss of structural memory leads to the reappearance of small clusters. 
These results provide a real-space structural interpretation of ionic 
transport in non-crystalline solids and support fractal pathway models of 
high-frequency ionic response.
\end{abstract}

\maketitle

The Universal Dielectric Response (UDR), also known as Jonscher’s law, 
provides a unifying description of the AC electrical response of a wide 
variety of materials exhibiting relaxation and conduction phenomena. 
Originally proposed by Jonscher in 1977~\cite{Jon1977}, this empirical 
law captures the frequency dependence of the AC conductivity $\sigma(\omega)$ 
observed in many dielectrics and semiconductors, particularly in 
structurally disordered systems. Its broad applicability suggests 
the existence of a common underlying charge-transport mechanism 
operating across diverse classes of materials.

For the real part of the complex conductivity, Jonscher’s law is 
commonly written as
\begin{equation}
\sigma'(\omega)=\sigma_{\text{DC}}+A\omega^n,
\label{Jonscher}
\end{equation}
where $\sigma_{\text{DC}}=\sigma'(\omega\to 0)$ is the DC conductivity, 
$A$ is a temperature-dependent prefactor, $\omega$ is the angular frequency, 
and $n$ is the Jonscher exponent, typically in the range $0<n\leq1$. 
The value of $n$ provides insight into the dominant transport mechanism: 
small values correspond to nearly frequency-independent conduction, 
while increasing $n$ reflects correlated hopping processes and stronger 
interactions between charge carriers and the surrounding matrix. 
In the limiting case $n=1$, the response approaches ideal Debye behaviour, 
characteristic of non-interacting dipoles with a single relaxation time.

The essential difference between Jonscher’s law and the classical Debye 
model lies in the complexity of the relaxation dynamics they describe. 
In disordered materials, charge carriers are localized in a broad 
distribution of energy wells and must hop between sites to contribute 
to transport. At low frequencies, carriers can explore long distances, 
giving rise to DC conduction. At higher frequencies, the rapidly oscillating 
field restricts motion to short-range, localized 
displacements~\cite{Bal2022,Roling1999}. While such motion contributes 
weakly to long-range transport, it strongly enhances dielectric losses, 
leading to the characteristic power-law increase of the AC 
conductivity~\cite{Sok2001,Bal2013,Mon2012}. Jonscher’s law effectively 
captures this correlated, multi-scale dynamics through an implicit 
distribution of relaxation times, in contrast to the single-time-scale 
dynamics assumed in the Debye model.

Ionic glasses represent a paradigmatic class of materials exhibiting 
Universal Dielectric Response. Their intrinsic structural disorder 
arises from a covalent glassy network within which mobile ions act as 
charge carriers. These ions experience a complex and heterogeneous energy 
landscape, characterized by a broad distribution of site energies and 
hopping barriers that are neither spatially uniform nor dynamically 
equivalent~\cite{Sok2001,Bal2013,Fre2013,Roling1998}. 
Despite extensive experimental and computational efforts, 
a complete microscopic understanding of ionic transport in these 
systems remains elusive.

A recurring concept in this context is that of ion-conducting pathways: 
preferential regions of enhanced mobility embedded within the disordered 
structure. Diffusive transport in non-crystalline systems has long been 
associated with such regions, which were anticipated as near-one-dimensional, 
fractal objects well before they could be directly resolved. Clarifying 
their geometry, temporal stability, and contribution to macroscopic 
transport remains a central challenge, requiring the analysis of ion 
dynamics across multiple spatial and temporal 
scales~\cite{Funke2015,Lee1991,Now1998,Haba2002}.

In this work, we investigate a lithium metasilicate glass from a 
microscopic perspective, focusing on the temporal evolution of 
lithium-ion dynamics in the presence of strong dynamic heterogeneity. 
Our primary objective is to identify and characterize the ion-conducting 
pathways explored by the most mobile ions, determine their geometrical 
properties, and establish a direct connection with the high-frequency 
regime of the AC ionic conductivity, corresponding to short times and 
limited spatial exploration. By linking microscopic ion dynamics to 
macroscopic dielectric response, this study provides new insight into the 
structural origin of transport anomalies in ion-conducting glasses and 
contributes to the rational design of non-crystalline solid-state electrolytes.

{\bf Numerical simulations.-}
To investigate the evolution of lithium-ion dynamics upon approaching 
the glass transition from the liquid state, we focus on lithium 
metasilicate (Li$_2$SiO$_3$), a model ion-conducting glass that has 
been extensively studied in the literature. Previous 
analyses~\cite{Ver2024,Ver2021} have reported pronounced temperature-dependent 
effects in this system, particularly in the mobility of lithium ions 
within the glassy network, providing valuable insight into transport 
mechanisms near the glass transition.

The system consists of 3456 atoms, corresponding to the Li$_2$SiO$_3$ 
stoichiometry, and is modelled using classical molecular dynamics simulations. 
Interatomic interactions are described by a Gilbert–Ida–type pair 
potential~\cite{Ida1976}, which includes long-range Coulomb interactions 
with effective charges, a dispersive $r^{-6}$ term acting between oxygen ions, 
and a short-range Born–Mayer repulsive contribution. 
The interaction parameters were taken from previous studies and 
validated for this system~\cite{Hab1992,Hab1993,Hab2002} through ab 
initio molecular orbital calculations.

Atomic velocities were initially assigned according to a Maxwell–Boltzmann 
distribution at $3500$ K. The system was first equilibrated at this 
temperature for $2$ ns in the microcanonical (NVE) ensemble. It was then 
cooled to the target temperatures of $700$ K and $1100$ K using an 
identical quench rate in all cases. Cooling was performed in two 
consecutive stages, each consisting of a $2$ ns simulation in the 
isothermal–isobaric (NPT) ensemble, during which a thermostat imposed a 
linear decrease in temperature while maintaining the pressure at $1$ atm. 
At each target temperature, an additional $2$ ns NPT equilibration was 
carried out to ensure thermal stability and to attenuate  pressure fluctuations 
typical of solid phases. The corresponding experimental densities were 
reproduced within reported uncertainties. From the temperature dependence of 
the system volume, the glass-transition temperature was identified 
as $T_g=1200$ K~\cite{Ver2021,Ver2024}.

The equations of motion were integrated using the Verlet algorithm with 
periodic boundary conditions and a time step of $1$ fs, as implemented in the 
LAMMPS simulation package~\cite{Pli1995}. Production trajectories were 
generated in the NPT ensemble. For subsequent analysis, 101 configurations 
were extracted from each trajectory at intervals of $200$ fs, ensuring 
statistical independence while avoiding significant structural relaxation effects~\cite{Bal2015}.

{\bf Autocorrelation function.-}
The velocity autocorrelation function  $Z(t)$ (VACF) 

\begin{equation}
	Z(t)=\frac{\sum_i |\mathbf{v}_i(t)\cdot\mathbf{v}_i(0)|^2} {\sum_i |\mathbf{v}_i(0)\cdot\mathbf{v}_i(0)|^2} 
\end{equation}
is the standard tool for linking microscopic dynamics with transport properties in disordered ionic systems.
It provides key information about the temporal evolution of ionic motion and is 
typically interpreted through four features: (i) an initial rapid decay 
associated with vibrational motion within the transient oxygen cage 
(femtosecond–picosecond scale); (ii) a negative minimum (backscattering), 
common in liquids and disordered solids, where an ion reverses direction after 
colliding with neighbours; (iii) a long-time decay to zero, signalling the 
onset of diffusive motion; and (iv) its direct relation to the diffusion 
coefficient via the Green–Kubo formula. 
Accordingly, the VACF is widely used to probe relaxation processes and
microscopic transport, including dynamical changes across the glass
transition~\cite{Erp1985,Fuk1981,Hok1984}.

To investigate dynamic heterogeneity, we focus on the lithium ions, the most 
mobile species in the system. A segmentation procedure is applied in which 
all Li atoms are first considered, and a mobility criterion is introduced: 
ions whose positional standard deviation exceeds $1\AA$ 
are classified as fast, whereas the remaining ions are 
labelled slow~\cite{Ver2024}. 

This procedure yields two mobility-resolved subsets at fixed temperature. 
The VACF is then computed separately for the fast and slow lithium ions 
and compared with the VACF averaged over all Li$^+$. 
Our working hypothesis is that fast ions should exhibit a more
rapid loss of velocity correlation, characterized by reduced backscattering
and a faster decay to zero, consistent with weaker confinement by the oxygen
network and a more liquid-like local environment.

\begin{figure}
       \includegraphics[width=\linewidth]{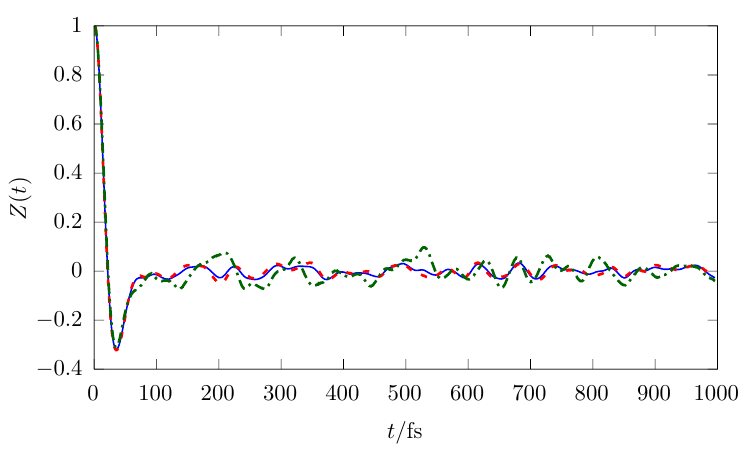}\\
       \includegraphics[width=\linewidth]{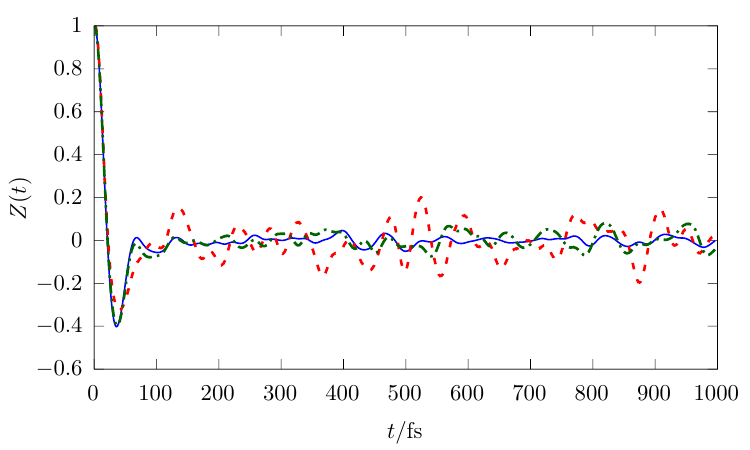}
	\caption{(color online) Lithium velocity autocorrelation function as a function 
	of time fot two	temperatures: 1000 K (top) and 700 K (bottom). In each
	panel, the 
	dashed (red) line corresponds to fast ions, the dash-dotted (green) 
	line to slow ions 
	and solid (blue) line to the ensemble average}
	\label{vacf}
\end{figure}

Figure~\ref{vacf} shows the VACFs computed for the full lithium ensemble and
for the two mobility subsets at temperatures below the glass transition.
In all cases, the VACF decorrelates within only a few timesteps, indicating
that ensemble-averaged velocity correlations are largely insensitive to the
pronounced dynamic heterogeneity reported previously~\cite{Bal2022,Ban2001}.
A more detailed picture emerges when the analysis is carried out at the
level of individual ions. In this case, the VACFs exhibit intermittent
episodes in which they approach values close to unity, signalling nearly
collinear motion over short time intervals. These events provide direct
evidence of transient, less-restricted displacements along preferred
directions, consistent with the existence of ion-conducting pathways.
Importantly, such episodes occur asynchronously across different ions and
are therefore strongly suppressed by ensemble averaging.
Furthermore, although the most mobile ions explore extended, three-dimensional 
trajectories, the directional correlations associated with these motions are 
effectively washed out when averaged over the full ensemble. As a consequence, 
the geometric signature of the conduction pathways remains hidden in 
conventional VACF analyses.
Taken together, these results show that ensemble-averaged VACFs, despite 
their widespread use, do not resolve the structural features of the 
diffusive pathways governing ionic transport in non-crystalline solids. 
This motivates the spatially resolved pathway analysis presented next.

{\bf Pathway structure.-}

Having established that ensemble-averaged velocity correlations are unable to 
resolve dynamic heterogeneity, while single-ion VACFs reveal short-lived 
episodes of nearly collinear motion, we now turn to the spatial origin of this 
behaviour. If highly mobile ions undergo transient intervals of directed 
motion, this must be reflected in the geometry and connectivity of the regions 
they explore. We therefore analyse the structure of the ion-conducting pathways,
with the aim of identifying the spatial organisation that supports these 
less-restricted displacements.

To this end, the simulation box is discretized into a cubic grid of 
approximately 200 sites per side. All sites are initially marked as inactive; 
a site becomes active as soon as it is visited by any lithium ion. As the 
simulation evolves, the number of active sites increases, forming clusters 
whose structure reflects the regions explored by ionic motion. Below the 
glass-transition temperature, the silicate network formed by silicon and 
oxygen atoms is highly rigid. As a result, the conduction pathways are 
expected to remain stable over time windows sufficiently long for the 
incoherent scattering function to remain close to unity~\cite{Heu2002,Kar2006}.

Lithium ions explore the system by performing random walks constrained by this 
network of conductive channels. Accordingly, the ion-conduction pathways 
can be inferred from the geometry of the clusters formed by active sites. 
Two limiting cases can be identified: ions confined to very small clusters 
remain localized near their initial positions, whereas ions belonging to large,
percolating clusters are able to contribute to long-range electrical 
conduction. More generally, each cluster contributes to the electrical 
response on characteristic length and time scales determined by its size 
and morphology.

\begin{figure}
       \includegraphics[width=\linewidth]{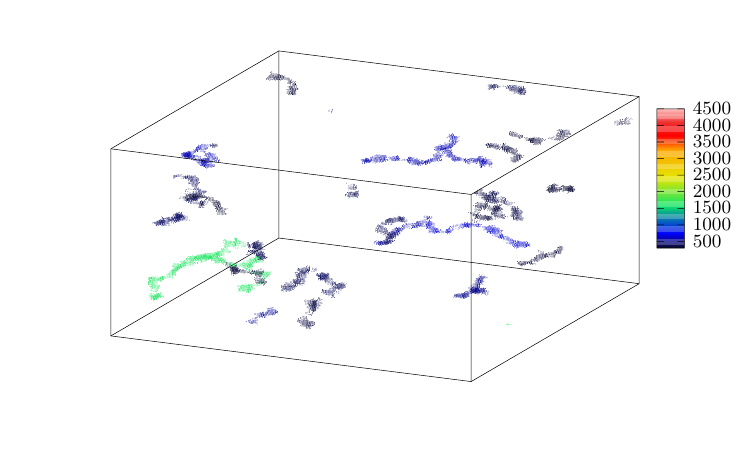}\\
       \includegraphics[width=\linewidth]{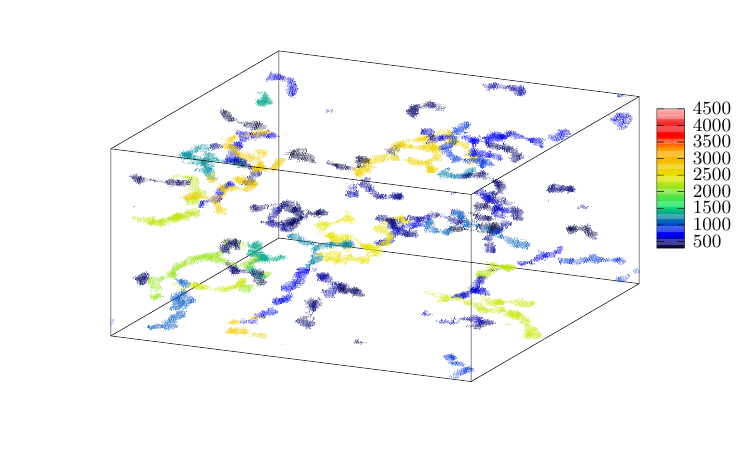}\\
        \includegraphics[width=\linewidth]{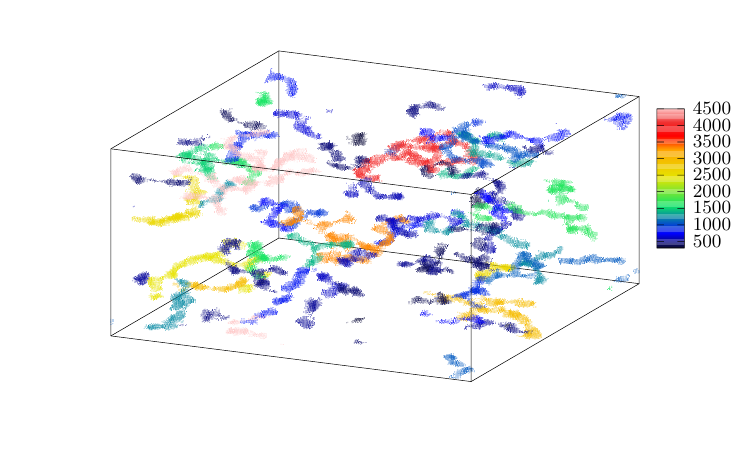}
	\caption{(color online) Clusters containing  more than $300$ active sites at 
	three different times. From top to bottom: $t=$ 5, 10, and 15 ps. 
	The system temperature is $T=700$ K. Cluster size is indicated by grey (color) scale}
       \label{clu_examples}
\end{figure}

To characterize the temporal evolution of the conduction pathways, we
systematically evaluated, at regular time intervals, the size $M$ of each
active-site cluster (number of active sites), its spatial extent $l$ (maximum
pairwise distance), and its box-counting fractal dimension $d_f$.
Figure~\ref{clu_examples} shows representative clusters containing more than
$300$ active sites for the system at $T=700$ K at three different times
($t=5$, $10$, and $15$ ps).

\begin{figure}
	\includegraphics[width=\linewidth]{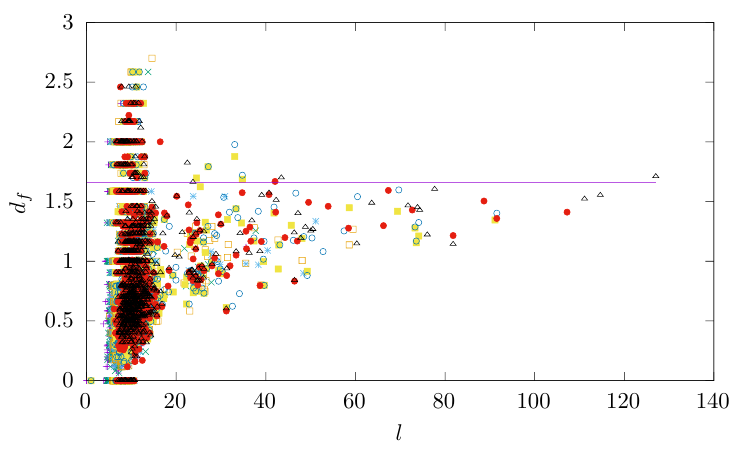}\\
	\includegraphics[width=\linewidth]{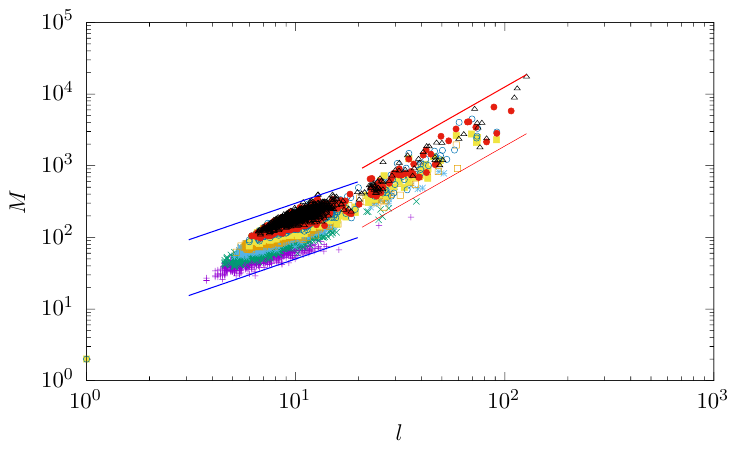}
	\caption{(color online) Cluster properties at $T=700$ K.\\
		Top panel:  box-counting fractal dimension as 
	a function of cluster length for all clusters at different times:  
	$t=$1 (violet plus), 2 (green cross), 4 (light-blue star), 
	6 (orange empty square), 10 (yellow filled square), 15 (blue empty-circle), 
	20 (red filled circle ), 30 (black empty triangle) ps. The horizontal line 
	indicates $d_f=1.7$. 
	Bottom panel: cluster size versus cluster length for the same times. The lines 
	with  slopes 1 (left) and 1.7 (right) are shown as guides to the eye.}
	\label{df_700}
\end{figure}

We first analised  the statistical properties of the conduction clusters at
$T=700$ K for times $t=1, 2, 4, 6, 10, 15, 20$ and $30$ ps. 
As shown prevously~\cite{Heu2002,Bal2014},
the incoherent scattering function remains close to unity over this time window,
indicating that the silicate backbone remains essentially stable.
The evolution of the fractal dimension as a function of cluster length is 
shown in Fig.~\ref{df_700} (top). At short times, most clusters are small 
and exhibit fractal dimensions close to $d_f\simeq 1$, indicating that, 
at small  spatial scales, the conducting pathways are quasi one-dimensional. 
As time progresses, some clusters grow and merge, forming larger structures, 
while others saturate and cease to evolve. The larger clusters 
display a nearly constant fractal dimension close to $d_f\simeq 1.7$, 
signalling the emergence of a branched, fractal geometry.

This behaviour is further illustrated in  Fig.~\ref{df_700} (bottom), where 
the cluster size $M$ is plotted against the cluster length $l$  on a log–log scale 
for the same times. Since these quantities are related 
through $M\sim l^{d_f}$, the data clearly show that, at short times, 
most clusters are nearly one-dimensional. At longer times, two distinct 
populations emerge: one consisting of small, quasi-one-dimensional clusters, 
and another corresponding to larger, branched clusters characterized by a 
fractal dimension close to $d_f\simeq 1.7$. Straight lines with slopes $1$ 
and $1.7$  are included as guides to the eye.

\begin{figure}
	\includegraphics[width=\linewidth]{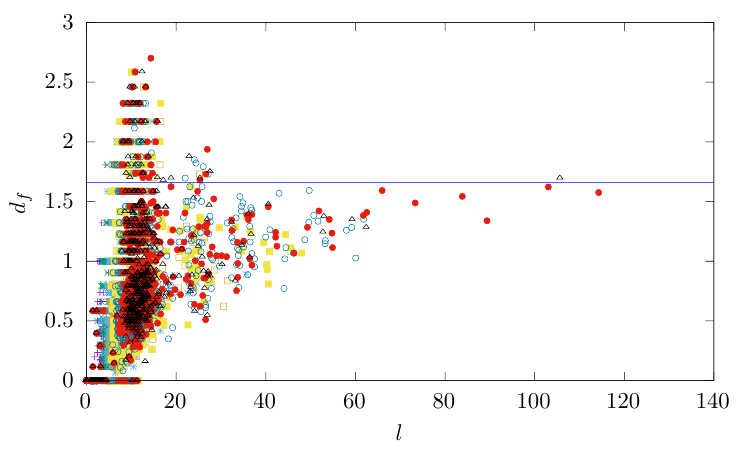}\\
	\includegraphics[width=\linewidth]{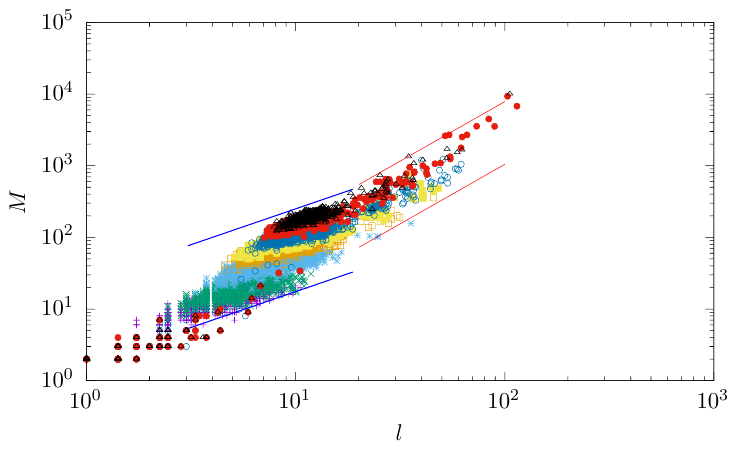}
	\caption{(color online) Cluster properties at $T=1100K$.\\
	Top panel:  box-counting fractal dimension as 
	a function of cluster length for times  $t=$ 0.10 (plus violet), 0.15 (cross green), 
	0.40 (star light-blue),	0.80 (empty-square orange), 1.20 (filled-square yellow), 
	2.50 (empty-circle blue), 
	5.00 (filled circle red), and 7.50 (empty triangle black) ps. 
	The horizontal line indicates $d_f=1.7$.
	Bottom panel; cluster size versus cluster length for the same times. Straight 
	lines with slopes slopes 1 (left) and 1.7 (right) are shown as guides to the eye. }
	\label{df_1100}
\end{figure}

To assess whether the ion-conducting pathways are ultimately determined by 
the rigid silicon–oxygen backbone that emerges at the glass transition, 
we also analysed the evolution of active-site clusters at $T=1100$ K, i.e., 
close to $T_g$. This comparison allows us to disentangle the role of 
structural arrest from purely dynamical effects in shaping the geometry 
of the conduction pathways.
Fig.~\ref{df_1100} shows, using the same analysis performed at $T=700$ K, 
the statistical properties of active-site clusters at $T=1100$ K and times $t=0.10, 
0.15, 0.40, 0.80, 1.20, 2.50, 5.00$  and $7.50$ ps. Although the time evolution is 
significantly much faster at this higher temperature, the results again indicate that 
at short times the conducting pathways are predominantly one-dimensional, while at longer 
times larger structures emerge, characterized by fractal dimension  $d_f\simeq 1.7$. 
Notably, at the longest times considered at $T=1100$ K, small clusters reappear. 
This behaviour indicates that the silicate network has begun to lose  memory of its initial 
condition; consistent with the pronounced decay  of the incoherent scattering function 
observed after a few picoseconds at this temperature~\cite{Heu2002,Bal2014}. 

{\bf Summary and conclusions.-}

We have investigated  ionic transport in a lithium metasilicate glass below 
and near the glass-transition temperature by combining mobility-resolved 
dynamics with a real-space characterization of ion-conducting pathways. 
Ensemble-averaged velocity autocorrelation functions rapidly decorrelate and 
fail to capture dynamic heterogeneity, whereas single-ion analysis 
reveals short-lived episodes of nearly collinear motion.
By mapping the regions explored by lithium ions, we show that ion-conducting 
pathways are quasi one-dimensional at short times, with a fractal 
dimension close to unity, and progresively evolve into larger branched 
structures characterized by a robust fractal dimension $d_f\simeq 1.7$. 
This geometry persists as long as the silicate backbone 
remains structurally arrested, while near $T_g$ the loss of structural 
memory leads to the reappearance of small clusters. 
Altogether these results provide microscopic evidence that 
ionic transport in non-crystalline solids proceeds through 
transient fractal pathways, offering a structural basis for the 
high-frequency regime of the Universal Dielectric Response observed 
in ion-conducting glasses.

{\bf Acknowledgements.-}
This work was supponted by the Universidad Nacional del Sur, PGI 24/Q146, 
the Universidad Nacional de Mar del Plata, 15/E1040, Agencia Nacional de 
Promoción Científica y Tecnológica (ANPCyT), PICT 2021-00288, and Consejo Nacional de 
Investigaciones Científicas y Técnicas (CONICET), PIP-21 GI11220200100317CO. The authors are Research Fellows at CONICET, Argentina.

\bibliography{My_Collection}{}
\bibliographystyle{apsrev4-1.bst}
\end{document}